\documentclass[iop]{emulateapj}
\usepackage{graphicx}
\usepackage{txfonts}
\usepackage{times}
\usepackage{url}
\usepackage{bm}
\usepackage{natbib}
\graphicspath{{./fig/}{./png/}}

\newcommand{\EQ}{\begin{equation}}
\newcommand{\EE}{\end{equation}}
\newcommand{\EQA}{\begin{eqnarray}}
\newcommand{\EEA}{\end{eqnarray}}

\newcommand{\pd}{\partial}

\newcommand{\DIV}{\bm{\nabla} \cdot }

\newcommand{\urms}{u_{\rm rms}}
\newcommand{\Urms}{U_{\rm rms}}

\newcommand{\kef}{k_{\rm f}}

\newcommand{\Rey}{{\rm Re}}
\newcommand{\Pe}{{\rm Pe}}
\newcommand{\Pra}{{\rm Pr}}
\newcommand{\Ta}{{\rm Ta}}
\newcommand{\Ra}{{\rm Ra}}

\newcommand{\Co}{{\rm Co}}

\def\onethird{{\textstyle{1\over3}}}
\def\onehalf{{\textstyle{1\over2}}}

\begin{document}

\title{Starspots due to large-scale vortices in rotating turbulent convection}

\author{Petri J.\ K\"apyl\"a$^{1,2}$, Maarit J.\ Mantere$^{1}$ and Thomas Hackman$^{1,3}$}
\affil{Physics Department, Gustaf H\"allstr\"omin katu 2a, PO Box 64,
FI-00014 University of Helsinki, Finland}
\affil{NORDITA, AlbaNova University Center, Roslagstullsbacken 23,
SE-10691 Stockholm, Sweden}
\affil{Finnish Centre for  Astronomy with ESO, University of Turku, 
V\"ais\"al\"antie 20, FI-21500 Piikki\"o, Finland}
\email{petri.kapyla@helsinki.fi
}

\begin{abstract} 
  We study the generation of large-scale vortices in rotating
  turbulent convection by means of Cartesian direct numerical
  simulations. We find that for sufficiently rapid rotation, cyclonic
  structures on a scale large in comparison to that of the convective
  eddies, emerge, provided that the fluid Reynolds number exceeds a
  critical value. For slower rotation, cool
  cyclonic vortices are
  preferred, whereas for rapid rotation, warm anti-cyclonic vortices
  are favored. In some runs in the intermediate regime both types of
  cyclones co-exist for thousands of convective turnover times.  The
  temperature contrast between the vortices and the surrounding
  atmosphere is of the order of five per cent. We relate the
  simulation results to observations of rapidly rotating late-type
  stars that are known to exhibit large high-latitude spots from
  Doppler imaging. In many cases, cool spots are accompanied with
  spotted regions with temperatures higher than the average.  In this
  paper, we investigate a scenario according to which the spots
  observed in the temperature maps could have a non-magnetic origin
  due to large-scale vortices in the convection zones of the stars.
\end{abstract} 

\keywords{Hydrodynamics -- convection -- turbulence}


\section{Introduction}
Rotating turbulent convection is considered to play a crucial role in
the generation of large-scale magnetic fields \citep{M78,KR80,RH04}
and differential rotation of stars \citep{R89}.  The interaction of
rotation and inhomogeneous turbulence leads to the so-called
$\alpha$-effect, which can sustain large-scale magnetic fields
\citep[e.g.][]{B01,KKB09b}. However, in many astrophysically relevant
cases large-scale shear flows are also present, which further
facilitate dynamo action by lowering the relevant critical dynamo
number.  In the Sun, for example, the entire convection zone is
rotating differentially \citep[cf.][]{Schouea98,Thompsonea03}, and a
meridional flow towards the poles is observed in the near surface
layers \citep*[e.g.][]{ZK04}. These flows are most often attributed to
rotationally influenced turbulent angular momentum and heat transport
\citep[cf.][]{R89,RC01,MBT06,KMGBC11}. In the solar case the
large-scale flows and also the magnetic activity are largely
axisymmetric \citep[e.g.][]{PBKT06}. This means that the sunspots,
which are concentrations of strong magnetic fields, are almost
uniformly distributed in longitude over the solar surface. The fact
that we observe the sunspots and can attribute magnetic fields to
them, has strongly influenced the interpretation of data from stars
other than the Sun.

The giant planets Jupiter and Saturn are also likely to have outer
convection zones \citep[e.g.][]{Busse76}, but they rotate much faster
than the Sun. Bands of slower and faster rotation alternate in their
atmospheres, reminiscent of rapidly rotating convection
\citep[e.g.][]{Busse94,HA07}. However, especially in Jupiter, large
spots in the form of immense storms are observed \citep{Marcus93}. 
Remarkably, the
largest of these, the Great Red Spot, has persisted at least 180
years. Similar features are observed also in Saturn
\citep[e.g.][]{saturn1991} and other giant planets.
The spots on giant planets are not of magnetic origin although dynamos
are likely to be present in the interiors of the planets. Thus their
explanation is probably related to hydrodynamical processes within the
convectively unstable layers.

Late-type stars with higher rotation velocities in comparison to the
Sun, on the other hand, often exhibit light curve variations that are
usually interpreted as large spots on the stellar surface
\citep[e.g.][]{Chuga66,Henry1995}.
In some cases the observational data
can be fitted with a model with two large spots at a 180 degree
separation in
longitude \citep{BT98}. There is also evidence that 
these `active longitudes' are not equal in strength
\citep[e.g.][]{Jyri11,Marjaana11}, and that the relative strength
of the spots
can, at least temporarily, reverse in a process dubbed 'flip-flop'
\citep[cf.][]{Jetsu1993}. One interpretation of the data is that the
spots are of magnetic origin and that the flip-flops are related to
magnetic cycles reminiscent of the solar cycle \citep[e.g.][]{BBIT98}.
On the other hand, it has been proposed that the flip-flops are only
short-term changes related to the activity cycle, while the structure
generating the temperature minima would migrate in the orbital
reference frame. This can be interpreted as an azimuthal dynamo wave
\citep[e.g.][]{Jyri11,Marjaana11}. Again, this interpretation relies
on the magnetic nature of the cool spots.

The cool spots detected by photometry and Doppler imaging using
spectroscopic observations have been taken as an indirect proxy of the
magnetic field on the stellar surface, deriving from the analogy to
sunspots -- strong magnetic field hinders convection and causes the
magnetized region to be cooler than its surrounding. Zeeman-Doppler
imaging of spectropolarimetric observations
\citep[e.g.][]{Semel89,Donati89,PK02,Carroll07} provides means to
directly measure the magnetic field strength and orientation on the
stellar surface. In the study of \citet{Donati97} spectropolarimetric
observations of several stars were collected during 23 nights
extending over a five year interval. They report that the Zeeman
signatures of cool stars almost always exhibit a very complex
shape with many successive sign reversals. This points to a rather
complicated field structure with different magnetic regions of
opposite polarities. Furthermore, these active regions were
mostly 500 to 1,000~K cooler than, and sometimes at the same
temperature as, but never warmer than the surrounding photosphere. In
the published temperature and magnetic field maps for AB Dor
\citep[][]{DC97}, however, no clear correlation between temperature
and magnetic field strength can be seen: in the temperature maps a
pronounced cool polar cap with weak fringes towards lower latitudes
are visible, whereas the strongest magnetic fields are seen as patchy
structures at lower latitudes with a clearly different distribution
in comparison to the temperature structures. Similar decorrelation of 
temperature
minima and magnetic field strength has been reported with the same
method for different objects \citep[e.g.][]{Donati99,Jeffers11}, and
also for the same objects with different methods
\citep[e.g.][]{Hussain2000,Oleg11}. The phenomenon, therefore, seems
to be wide-spread, and method-independent.

One possible explanation to the decorrelation of magnetic field and
temperature structures could 
be that there is simply less light coming from the spotted parts
than from the unspotted surface. Thus the Zeeman signatures from cool spots
may be ``drowned'' in the signal from the unspotted surface or bright 
features. However, this should lead to systematic effects where the 
detected magnetic field strength would be correlated with the surface 
temperature.
The least-squares deconvolution technique \citep[e.g.][]{Donati97}, 
which is necessary for enhancing the Zeeman signal,
may influence the temperature and magnetic Doppler imaging
differently. The
latitudes of any surface features in Doppler images are always more unreliable
than the longitudes, a fact that will not make a comparison of temperature and 
magnetic field maps any easier. One could thus expect, that there
could be artificial discrepancies in the latitudes of magnetic and
temperature features. Still, the lack of connection between even the
longitudes of cool spots and magnetic features is surprising.

In this paper we consider a completely different scenario, according
to which the formation of temperature anomalies on the surfaces of
rapidly rotating late-type stars could occur due to a hydrodynamical
instability generating large-scale vortices, analogously to the giant
planets in the solar system. To manifest this mechanism in action, we
simulate rotating turbulent convection in local Cartesian domains,
representing parts of the stratified stellar convection zones located
near the polar regions. We show that under such a setting, large-scale
vortices or cyclones are indeed generated provided that the rotation
is sufficiently rapid and the Reynolds number exceeds a critical
value. Depending on the handedness of the vortex, which on the other
hand depends on the rotation rate, the resulting spot can be cooler or
warmer than the surrounding atmosphere.

We acknowledge that our model is rather primitive, lacking realistic
radiation transport, spherical geometry, and relying on a polytropic
setup for the 
stratification. Therefore, a
detailed comparison with observations
is not possible at this point. However, the main purpose of the
present paper is to provide a proof of concept of the existence of
large-scale vortices with temperature anomalies close to those
observed in rapidly rotating hydrodynamic convection.  We also note
that similar large-scale cyclonic structures have recently been
reported from large-eddy simulations of turbulent convection
\citep{Chan03,Chan07}. We make comparisons to these studies when
possible.

\section{The model}
\label{sec:model}
Our model setup is the same as that used by \cite{KKB09b} but without 
magnetic fields. A
rectangular portion of a star is modeled by a box situated at
colatitude $\theta$. The box is divided into three layers: an upper
cooling layer, a convectively unstable layer, and a stable overshoot
layer (see below). We solve the following set of equations for
compressible hydrodynamics:
\begin{equation}
\frac{\mathrm{D} \ln \rho}{\mathrm{D}t} = -\DIV{\bm U},
 \end{equation}
\begin{equation}
 \frac{\mathrm{D} \bm U}{\mathrm{D}t} = -\frac{1}{\rho}{\bm \nabla}p + {\bm g} - 2\bm{\Omega} \times \bm{U} + \frac{1}{\rho} \bm{\nabla} \cdot 2 \nu \rho \mbox{\boldmath ${\sf S}$}, \label{equ:UU}
 \end{equation}
\begin{equation}
 \frac{\mathrm{D} e}{\mathrm{D}t} = - \frac{p}{\rho}\DIV {\bm U} + \frac{1}{\rho} \bm{\nabla} \cdot K \bm{\nabla}T + 2 \nu \mbox{\boldmath ${\sf S}$}^2 - \frac{e\!-\!e_0}{\tau(z)}, \label{equ:ene}
 \end{equation}
where $\mathrm{D}/\mathrm{D}t = \pd/\pd t + \bm{U} \cdot \bm{\nabla}$ 
is the advective time derivative,
$\nu$ is the kinematic viscosity, $K$ is the heat 
conductivity, $\rho$ is the density, $\bm{U}$ is the
velocity, $\bm{g} = -g\hat{\bm{z}}$ is the gravitational acceleration,
and $\bm{\Omega}=\Omega_0(-\sin \theta,0,\cos \theta)$ is the rotation vector.
The fluid obeys an ideal gas law $p=(\gamma-1)\rho e$, where $p$
and $e$ are the pressure and the internal energy, respectively, and
$\gamma = c_{\rm P}/c_{\rm V} = 5/3$ is the ratio of specific heats at
constant pressure and volume, respectively.
The specific internal energy per unit mass is related to the
temperature via $e=c_{\rm V} T$.
The rate of strain tensor $\mbox{\boldmath ${\sf S}$}$ is given by
\begin{equation}
{\sf S}_{ij} = \onehalf (U_{i,j}+U_{j,i}) - \onethird \delta_{ij} \DIV \bm{U}.
\end{equation}
The last term of Eq.~(\ref{equ:ene}) describes cooling at the top of
the domain. 
Here $\tau(z)$ is a cooling time which has a profile
smoothly connecting the upper cooling layer and the convectively
unstable layer below, where $\tau\to\infty$.

The positions of the bottom of the box, bottom and top of the
convectively unstable layer, and the top of the box, respectively,
are given by $(z_1, z_2, z_3, z_4) = (-0.85, 0, 1, 1.15)d$, where $d$ is 
the depth of the convectively unstable layer. Initially
the stratification is piecewise polytropic with polytropic indices
$(m_1, m_2, m_3) = (3, 1, 1)$, which leads to a convectively unstable
layer above a stable layer at the bottom of the domain. 
In a system set up this way, convection transports roughly 20 per 
cent of the 
total flux \citep[cf.][]{BCNS05}.
Due to the presence of the cooling term, a stably stratified
isothermal layer forms at the top. The horizontal extent of the
box, $L_{\rm H}\equiv L_x=L_y$, is $4d$.  
All simulations with rotation
are made at the North pole, corresponding to $\theta=0\degr$. 
The simulations were performed with the {\sc Pencil
  Code}\footnote{http://code.google.com/p/pencil-code/}, which is a
high-order finite difference method for solving the compressible
equations of magnetohydrodynamics.

\subsection{Units and non-dimensional parameters}
Non-dimensional quantities are obtained by setting
\begin{eqnarray}
d = g = \rho_0 = c_{\rm P} = 1\;,
\end{eqnarray}
where $\rho_0$ is the initial density at $z_2$. The units of length, time,
velocity, density, and entropy are
\begin{eqnarray}
&& [x] = d\;,\;\; [t] = \sqrt{d/g}\;,\;\; [U]=\sqrt{dg}\;,\;\; \nonumber \\ && [\rho]=\rho_0\;,\;\; [s]=c_{\rm P}.
\end{eqnarray}
We define the Prandtl number and the Rayleigh
number as
\begin{eqnarray}
\Pra=\frac{\nu}{\chi_0}\;,\;\; \Ra=\frac{gd^4}{\nu \chi_0} \bigg(-\frac{1}{c_{\rm P}}\frac{{\rm d}s}{{\rm d}z
} \bigg)_0\;,
\end{eqnarray}
where $\chi_0 = K/(\rho_{\rm m} c_{\rm P})$ is the thermal
diffusivity, and $\rho_{\rm m}$ is the density in the middle of
the unstable layer, $z_{\rm m} = \onehalf(z_3-z_2)$. The entropy 
gradient, measured at $z_{\rm m}$, in the non-convective hydrostatic state,
is given by
\begin{eqnarray}
\bigg(-\frac{1}{c_{\rm P}}\frac{{\rm d}s}{{\rm d}z}\bigg)_0 = \frac{\nabla-\nabla_{\rm ad}}{H_{\rm P}}\;,
\end{eqnarray}
where $\nabla-\nabla_{\rm ad}$
is the superadiabatic temperature gradient with 
$\nabla_{\rm  ad} = 1-1/\gamma$, $\nabla = (\pd \ln T/\pd \ln
  p)_{z_{\rm m}}$, and where $H_{\rm P}$ is the pressure scale height.
The amount of stratification is determined by the parameter 
$\xi_0 =(\gamma-1) e_0/(gd)$, which is the pressure scale height at
the top of the domain normalized by the depth of the unstable layer.
We use $\xi_0 =1/3$ in all cases,
which results in a density contrast of about 23 across the domain.
We define the Reynolds and P\'eclet numbers via
\begin{eqnarray}
{\rm Re} = \frac{\urms}{\nu \kef}\;,\;\; {\Pe} = \frac{\urms}{\chi_0 \kef} = \Pr\ {\rm Re}\;,
\end{eqnarray}
where $\kef = 2\pi/d$ is adopted as an estimate
for the wavenumber of the energy-carrying eddies, and
$\urms=\sqrt{3 u_z^2}$. This definition of $\urms$ neglects the 
contributions from
the large-scale vortices that are generated in the rapid
rotation regime.
Note that with our definitions $\Rey$ and $\Pe$ are smaller than the
usual one by a factor of $2\pi$.
The amount of rotation is quantified by
the Coriolis number, defined as
\begin{eqnarray}
{\rm Co} = \frac{2\Omega_0}{\urms \kef}\;. \label{equ:Co}
\end{eqnarray}
We also quote the value of the Taylor number,
\begin{equation}
\Ta=\left(2\Omega_0 d^2/\nu\right)^2,
\end{equation}
which is related to the Ekman number via ${\rm Ek}=\Ta^{-1/2}$.

\subsection{Boundary conditions}

The horizontal boundaries are periodic for all variables. Stress-free
conditions are used for the velocity
at the vertical boundaries.
\begin{eqnarray}
U_{x,z}=U_{y,z}=U_z=0.
\end{eqnarray}
Temperature is kept constant on the upper boundary and the temperature
gradient
\begin{eqnarray}
\frac{dT}{dz}=\frac{-g}{c_{\rm V}(\gamma-1)(m+1)},
\end{eqnarray}
is held constant at the lower boundary, yielding a constant heat flux
$F_0=-K \pd T/\pd z$ through the lower boundary.

\begin{deluxetable*}{cccccccccccc}
\tabletypesize{\scriptsize}
\tablecaption{Summary of the runs. 
  Here, $\mbox{\rm Ma}=\Urms/(gd)^{1/2}$ and 
  $\mbox{\rm Ma}_{\rm z}=\urms/(gd)^{1/2}$. Brackets indicate that 
  the simulation has not been run to a saturated state. The dimensionless input 
  heat flux at the lower boundary of the box is given by 
  $\tilde{F}_0=F_0/(\rho c_{\rm s}^3)$, where $c_{\rm s}$ is the 
  adiabatic sound speed and $\rho$ is the density, both measured at the lower 
  boundary of the domain. The last column indicates the 
  presence of cyclonic (C), anti-cyclonic (A), or both types (A+C) of 
  vortices.}
\tablewidth{0pt}
\tablehead{
\colhead{Run} & \colhead{grid} & \colhead{$\mbox{Ma}$} & 
\colhead{${\rm Ma}_{\rm z}$} & \colhead{$\Rey$} & \colhead{$\Pe$} &
\colhead{$\Pra$} & \colhead{$\Ra$} & \colhead{$\Co$} & 
\colhead{$\mbox{Ta}$} & \colhead{$\tilde{F}_0$} & \colhead{Cyclones}
}
\startdata
A1   & $256^2\times 128$  &  $0.048$  &  $0.020$  & $33$ & $8$ & $0.24$  & $2.0\cdot10^6$ & $15.5$ & $4.0\cdot10^8$   & $1.7\cdot10^{-5}$ & yes (A) \\ 
A2   & $256^2\times 128$  &  $0.018$  &  $0.017$  & $13$ & $6$ & $0.48$  & $1.0\cdot10^6$ & $14.4$ & $5.6\cdot10^7$ & $1.7\cdot10^{-5}$ & no  \\ 
A3   & $256^2\times 128$  &  $0.022$  &  $0.019$  & $21$ & $7$ & $0.36$  & $1.3\cdot10^6$ & $12.3$ & $1.0\cdot10^8$ & $1.7\cdot10^{-5}$ & no  \\ 
A4   & $256^2\times 128$  & $(0.063)$ &  $0.023$  & $37$ & $9$ & $0.24$  & $2.0\cdot10^6$ & $10.3$ & $2.3\cdot10^8$ & $1.7\cdot10^{-5}$ & yes (A) \\ 
A5   & $256^2\times 128$  &  $0.021$  &  $0.020$  & $16$ & $9$ & $0.48$  & $1.0\cdot10^6$ & $7.9$ & $2.5\cdot10^7$ & $1.7\cdot10^{-5}$ & no  \\ 
A6   & $256^2\times 128$  &  $0.024$  &  $0.023$  & $24$ & $9$ & $0.36$  & $1.3\cdot10^6$ & $7.0$ & $4.4\cdot10^7$ & $1.7\cdot10^{-5}$ & no  \\ 
A7   & $256^2\times 128$  & $(0.093)$ &  $0.026$  & $42$ & $10$ & $0.24$  & $2.0\cdot10^6$ & $6.1$ & $1.0\cdot10^8$ & $1.7\cdot10^{-5}$ & yes (A+C) \\ 
A8   & $256^2\times 128$  &  $0.028$  &  $0.027$  & $28$ & $11$ & $0.36$  & $1.3\cdot10^6$ & $3.6$ & $1.6\cdot10^7$ & $1.7\cdot10^{-5}$ & no  \\ 
A9   & $256^2\times 128$  &  $0.082$  &  $0.028$  & $45$ & $11$ & $0.24$  & $2.0\cdot10^6$ & $3.4$ & $3.6\cdot10^7$ & $1.7\cdot10^{-5}$ & yes (C) \\ 
A9b  & $256^2\times 128$  &  $(0.070)$  &  $(0.031)$  & $49$ & $12$ & $0.24$  & $2.0\cdot10^6$ & $2.1$ & $1.6\cdot10^7$ & $1.7\cdot10^{-5}$ & decay \\ 
A10  & $256^2\times 128$  &  $0.032$  &  $0.033$  & $53$ & $13$ & $0.24$  & $2.0\cdot10^6$ & $1.0$ & $4.0\cdot10^6$ & $1.7\cdot10^{-5}$ & no  \\ 
A11  & $256^2\times 128$  &  $0.038$  &  $0.038$  & $61$ & $15$ & $0.24$  & $2.0\cdot10^6$ & 0 & $0$ & $1.7\cdot10^{-5}$ & no  \\ 
\hline
B1   & $256^2\times 128$  & $0.017$ &  $0.016$  & $26$ & $13$ & $0.48$  & $4.0\cdot10^6$ & $9.7$ & $1.0\cdot10^8$ & $8.6\cdot10^{-6}$ & no  \\ 
B2   & $256^2\times 128$  & $(0.021)$ & $(0.017)$ & $37$ & $13$ & $0.36$  & $5.4\cdot10^6$ & $9.1$ & $1.8\cdot10^8$ & $8.6\cdot10^{-6}$ & yes (A+C) \\ 
B3   & $256^2\times 128$  & $(0.034)$ & $(0.020)$  & $63$ & $15$ & $0.24$  & $8.0\cdot10^6$ & $8.0$ & $4.0\cdot10^8$ & $8.6\cdot10^{-6}$ &  yes (A+C) \\ 
\hline
C1   & $256^2\times 128$  & $0.011$ &  $0.011$  & $17$ & $16$ & $0.96$  & $8.0\cdot10^6$ & $14.8$ & $1.0\cdot10^8$ & $4.3\cdot10^{-6}$ & no  \\ 
C2   & $256^2\times 128$  & $(0.014)$ &  $(0.012)$  & $25$ & $18$ & $0.72$  & $1.1\cdot10^7$ & $13.6$ & $1.8\cdot10^8$ & $4.3\cdot10^{-6}$ & no  \\ 
C3   & $256^2\times 128$  & $(0.022)$ &  $(0.014)$  & $44$ & $21$ & $0.48$  & $1.6\cdot10^7$ & $11.6$ & $4.0\cdot10^8$ & $4.3\cdot10^{-6}$ & yes (A) \\ 
\hline
D1   & $256^2\times 128$  & $0.013$ &  $0.013$  & $42$ & $51$ & $1.20$  & $4.0\cdot10^7$ & $7.2$ & $1.4\cdot10^8$ & $1.7\cdot10^{-6}$ & no  \\ 
D2   & $512^2\times 256$  & $(0.038)$ &  $(0.013)$  & $101$ & $49$ & $0.48$  & $1.0\cdot10^8$ & $7.5$ & $9.0\cdot10^8$ & $1.7\cdot10^{-6}$ & yes (A+C) 
\enddata
\label{tab:runs}
\end{deluxetable*}

\section{Results}
\label{sec:results}
We perform a number of numerical experiments in order to determine the
conditions under which large-scale cyclones are excited.  The basic
input parameters and some key diagnostic outputs of the simulations
are listed in Table~\ref{tab:runs}. We perform a few (Set~A) or a
single (Sets~B, C, and D) progenitor run with a given input heat 
flux and approximately constant P\'eclet number in
each Set from which the rest of the runs are obtained by continuing
from a thermally saturated snapshot and changing the value of 
the kinematic
viscosity $\nu$ in order to change $\Rey$. The higher resolution run
D2 was remeshed from a lower resolution case D1.

\subsection{Excitation of large-scale vortices}

\begin{figure}[t]
\centering
\includegraphics[width=\columnwidth]{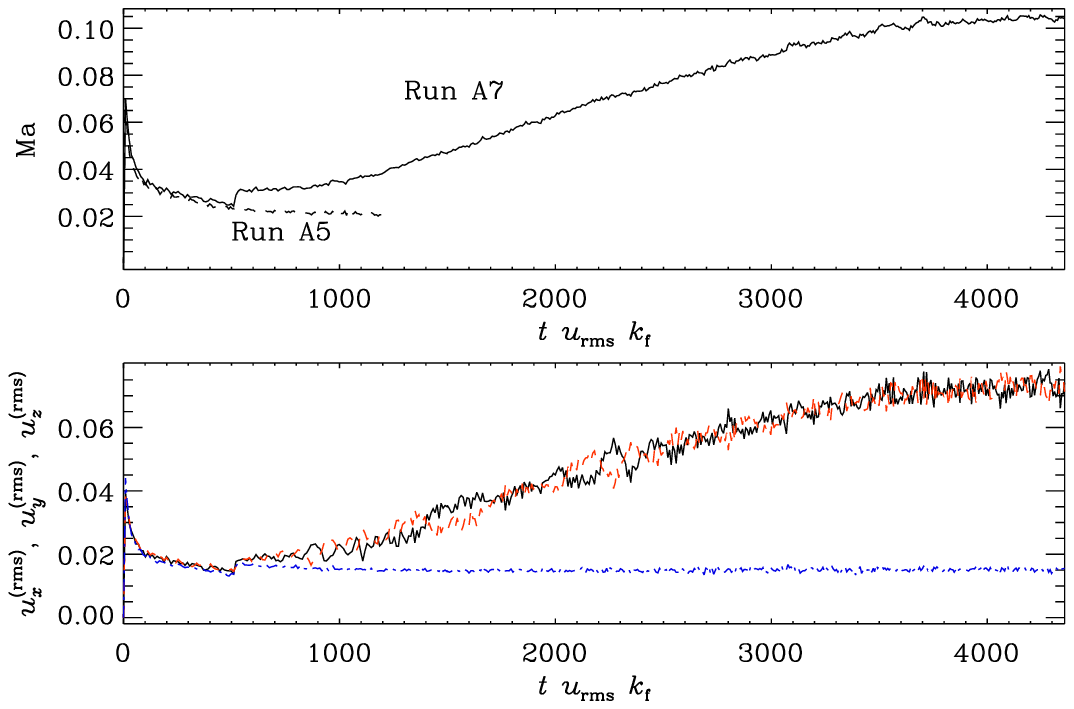}
\caption{Upper panel: total rms-velocity ${\rm Ma}=\Urms/(dg)^{1/2}$
  from Runs~A5 and A7. Lower panel: velocity components $\sqrt{u_x^2}$
  (black solid line), $\sqrt{u_y^2}$ (red dashed), and $\sqrt{u_z^2}$
  (blue dot-dashed) from Run~A7 in units of $(dg)^{1/2}$.  The jump at
  $t \urms \kef \approx 500$ is due to a lowering of $\nu$ at this
  point.}
\label{purms.eps}
\end{figure}

We perform several sets of runs where the P\'eclet number and input
energy flux are constant, whereas the Reynolds and Coriolis numbers
are varied. We are limited to exploring a small number of cases due to
the slow growth of the vortices, see Table~\ref{tab:runs}. Typically
the time needed for the saturation of the cyclones is several thousand
convective turnover times (see Fig.~\ref{purms.eps}).  For Run~A7 in
Fig.~\ref{purms.eps}, $t \urms \kef\approx4300$ corresponds to roughly
$9\tau_{\rm ther}$ or $2.2\cdot10^4\tau_{\rm dyn}$, where $\tau_{\rm
  ther}=d^2/\chi_0$ is the thermal diffusion time and $\tau_{\rm
  dyn}=\sqrt{d/g}$ is the dynamical free fall time. However, when the
input flux is lowered, by decreasing the heat conductivity, the
thermal relaxation time increases. For Runs~D1 and D2 the thermal
diffusion time is ten times longer than for runs in Set~A.  Thus many
of our runs were continued only until the presence or the absence of
the cyclones was apparent.

We find that a reliable diagnostic indicating the presence of
large-scale vortices is to compare the rms-value of the total
velocity, $\Urms$, and the volume average of the quantity
$\urms=\sqrt{3u_z^2}$. The latter neglects the horizontal velocity
components, which grow significantly
when large-scale cyclones are
present (see the lower panel of Fig.~\ref{purms.eps}). 
In the cyclone-free regime, irrespective of the rotation rate, we find
that $\Urms\approx\urms$ suggesting that the flow is only weakly
anisotropic (see Table~\ref{tab:runs}). In the growth phase of the
vortices one of the horizontal velocity components is always stronger,
but the relative strength of the components changes as a function
of time (see the lower panel of Fig.~\ref{purms.eps}). This undulation
is related to quasi-periodic changes of the large-scale pattern of the
flow, although their ultimate cause is not clear.

\begin{figure}[t]
\centering
\includegraphics[width=\columnwidth]{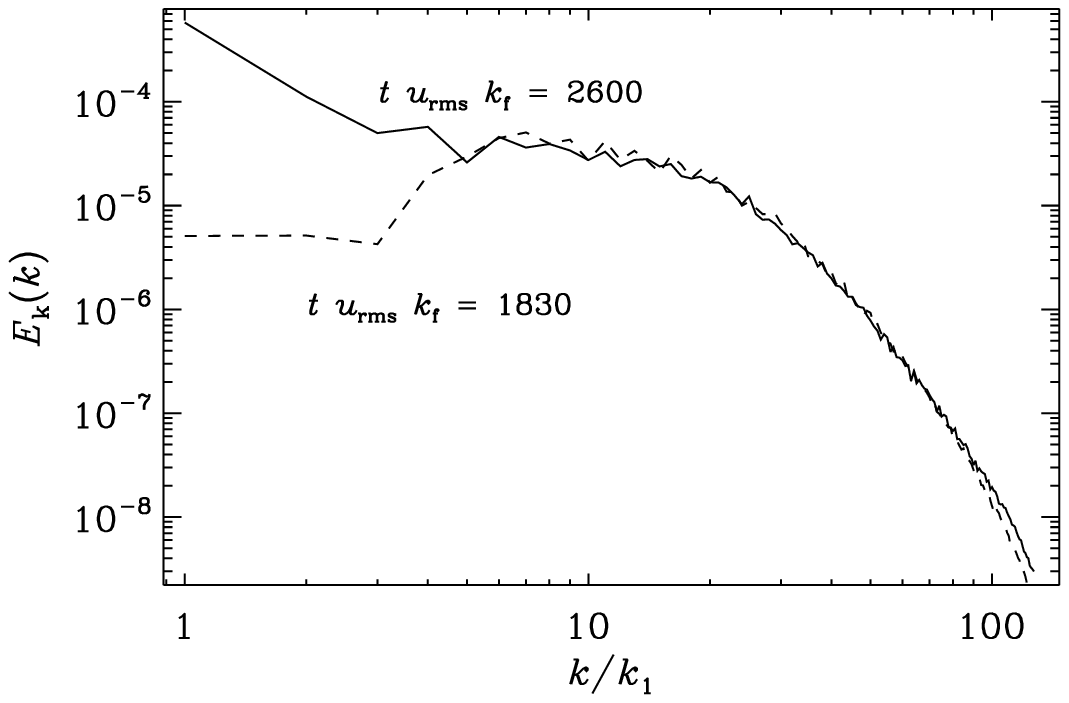}
\caption{Power spectra of velocity from early (dashed line) and late
  (solid) times from Run~B3 from $z=z_{\rm m}$.}
\label{pspec_256x128b1}
\end{figure}

\begin{figure}[t]
\centering
\includegraphics[width=\columnwidth]{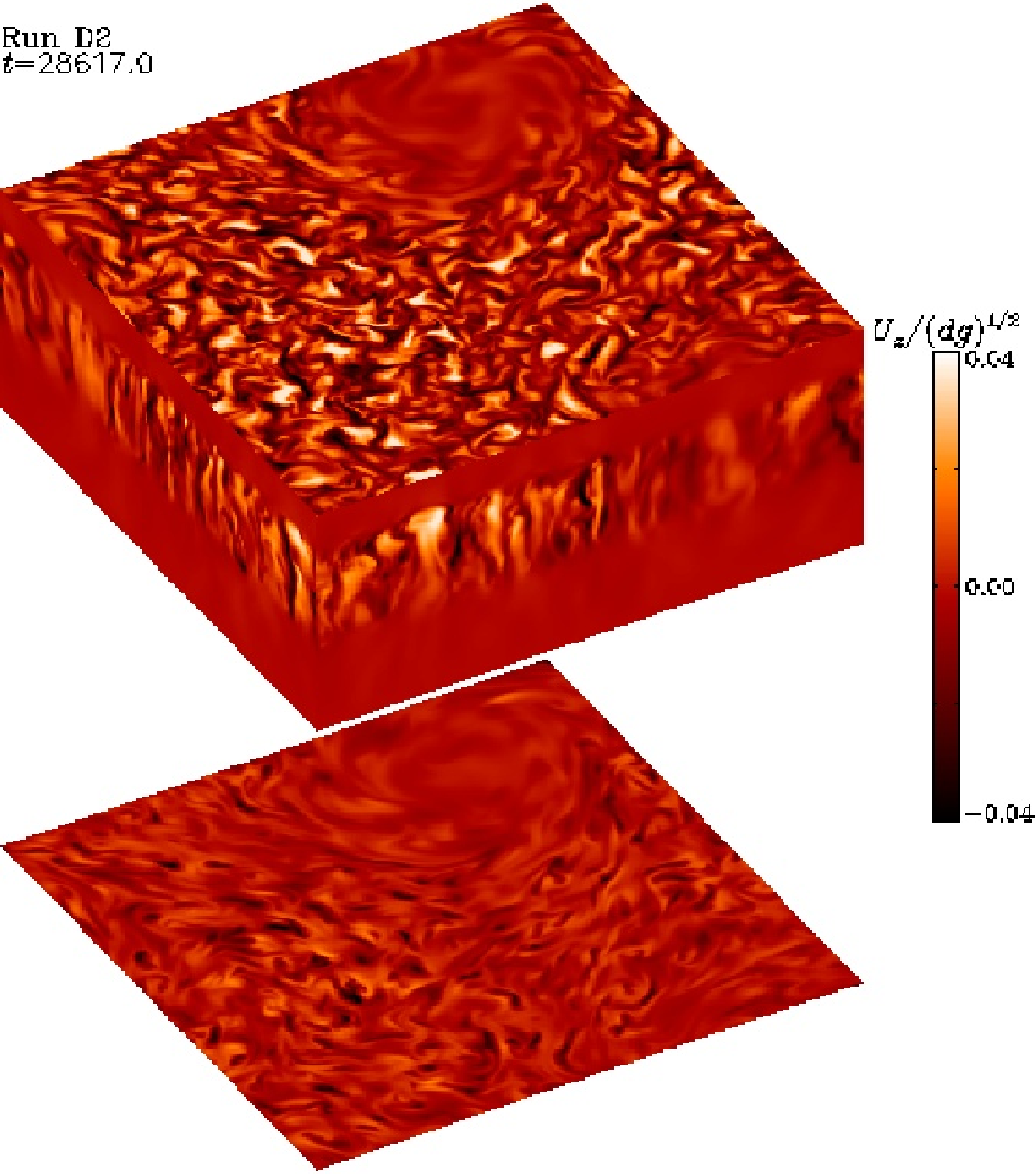}
\caption{Vertical velocity component $U_z$ at the periphery of the box
  from Run~D2. See also
  http://www.helsinki.fi/\ensuremath{\sim}kapyla/movies.html. The top
  and bottom panels show slices near the top and bottom of the
  convectively unstable layer, respectively.}
\label{psnap_512x256a2}
\end{figure}

Another quantitative diagnostic is to
monitor the power spectrum of the flow from a horizontal plane within
the convection zone. A typical example is shown in
Fig.~\ref{pspec_256x128b1} where power spectra of the velocity from
the middle of the convection zone at two different times from Run~B3 are
shown.
The snapshot at $t \urms \kef = 1830$ is the initial state for
Run~B3, taken from a lower Reynolds number Run~B1, showing no 
cyclones. The power spectrum
shows a maximum at $k/k_1=7$, indicating that most of the energy is
contained in structures having a size typical of convective
eddies. However, as the run is continued further, a large-scale
contribution due to the appearance of the vortices, peaking at
$k/k_1=1$ grows, and ultimately dominates the power spectrum.
We note that this run was not 
continued
until saturation so the peak at
$k/k_1=1$ is likely to be even higher in the final state. The
presence of the vortices is also clear by visual inspection of the
flow. A typical example is shown in Fig.~\ref{psnap_512x256a2}, where
the vertical velocity component, $U_z$, is shown from the periphery of
the domain for Run~D2.

The data in Table~\ref{tab:runs} suggests that large-scale vortices 
are excited provided the Reynolds number exceeds a critical value,
$\Rey_{\rm c}$. For $\Pe\approx 10$ (Set~A) we find that $\Rey_{\rm c}$
is around 30, although the sparse coverage of the parameter range does
not allow a very precise estimate to be made. We find a similar value
for $\Rey_{\rm c}$ in Sets~B and C, whereas for $\Pe\approx50$ in
Set~D, the critical Reynolds number is greater than 42.
In Set~C, Runs~C1 and C2 were started from a snapshot of Run~C3 at a
time when vortices were already clearly developing. In both cases we
find that the cyclones decay, suggesting that their presence is not
strongly dependent on the history of the run.

\begin{figure*}[t]
\centering
\includegraphics[width=0.8\textwidth]{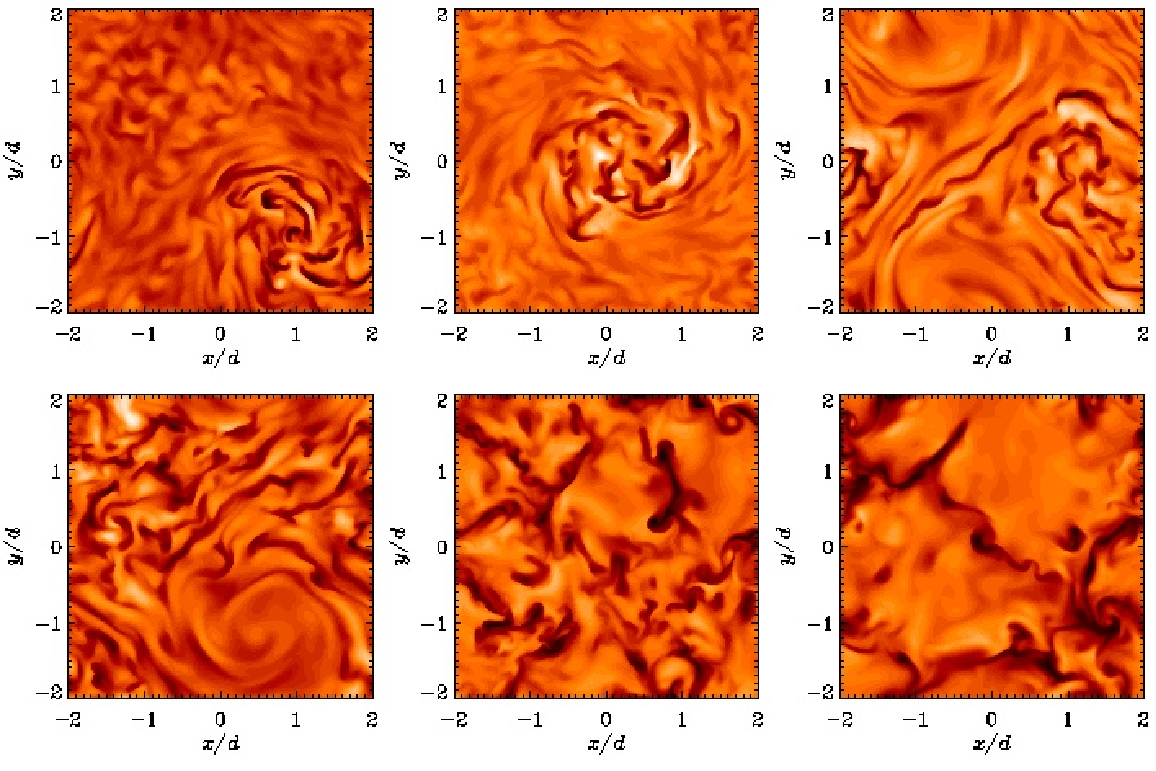}
\caption{From left to right: vertical velocity $U_z$ from $z=z_{\rm m}$ for
  Runs~A1, A4, A7 (upper row), A9, A10, and A11 (lower row).
  The rotational influence is decreasing from left to right and top 
  to bottom.}
\label{puu_slices}
\end{figure*}

The critical Coriolis number in Set~A is somewhere between 2.1 and
3.4. Again a very precise determination cannot be made, but continuing
from a saturated snapshot of Run~A9 with a somewhat lower rotation
rate indicates that the vortices decay (Run~A9b).
We have limited the present study to the North pole ($\theta=0$), but
vortices are also excited at least down to latitude $\theta=45\degr$
\citep[cf.][]{Chan07}.

\subsection{Thermal properties of the cyclones}

\begin{figure*}[t]
\centering
\includegraphics[width=0.8\textwidth]{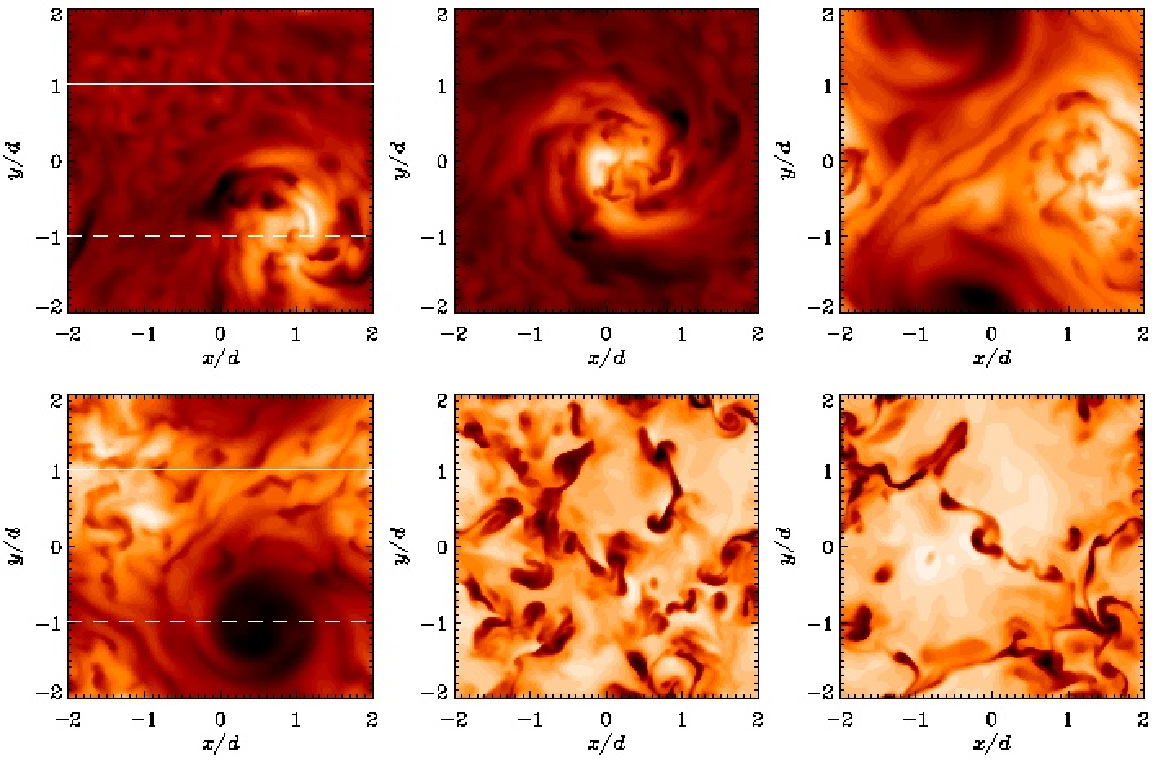}
\caption{From left to right: temperature from $z_{\rm m}$ for Runs~A1,
  A4, A7 (upper row), A9, A10, and A11 (lower row). The rotational 
  influence is decreasing from left to right and top to bottom. The
  solid and dashed horizontal lines on the leftmost panels 
  correspond to line
  plots shown in Fig.~\ref{ptempc}.}
\label{pT_slices}
\end{figure*}

In order to study the possible observable and other effects of the
vortices, we ran a few simulations in Set~A (Runs~A1, A4, A7, A9, A10,
and A11) to full saturation.  Figures~\ref{puu_slices} and
\ref{pT_slices} show the vertical velocity and temperature in the
saturated regime from the six runs listed above. In the non- and
slowly rotating cases (the two rightmost panels on the lower rows of
Figs~\ref{puu_slices} and \ref{pT_slices}), convection shows a typical
cellular pattern. Vorticity is generated at small scales at the
vertices of the convection cells, but no large-scale pattern arises.
We note that long-lived large-scale circulation can also emerge in
non-rotating convection \citep[e.g.][]{Bukai09}. However, such
structures are not likely to be of relevance in rapidly rotating
stars.

When the rotation is increased to $\Co\approx3.4$, a cyclonic, i.e.\
rotating in the same sense as the overall rotation of the star, vortex
appears (the lower left panels of Figs.~\ref{puu_slices} and
\ref{pT_slices}). Vertical motions are suppressed within the vortex
and it appears as a cool spot in the temperature slice. Increasing
rotation further to $\Co\approx6$, also an anti-cyclonic,
i.e.\ rotating opposite to the overall fluid rotation, warm 
vortex appears
(the rightmost upper panels of Figs.~\ref{puu_slices} and
\ref{pT_slices}). In Run~A7 the two vortices coexist for thousands of
convective turnover times. In the most rapidly rotating cases A1 and
A4 (the two leftmost panels in the upper rows of
Figs.~\ref{puu_slices} and \ref{pT_slices}) a single anti-cyclonic
vortex persists in the saturated regime. A similar behaviour as a
function of rotation was found by \cite{Chan07} from large-eddy
simulations. The anti-cyclonic vortices show vigorous convection
whereas in the surrounding regions convection appears suppressed.
Due to the enhanced energy transport by convection, the anti-cyclones
appear as warmer structures than their surroundings in the temperature
slices.

\begin{figure}[t]
\centering
\includegraphics[width=\columnwidth]{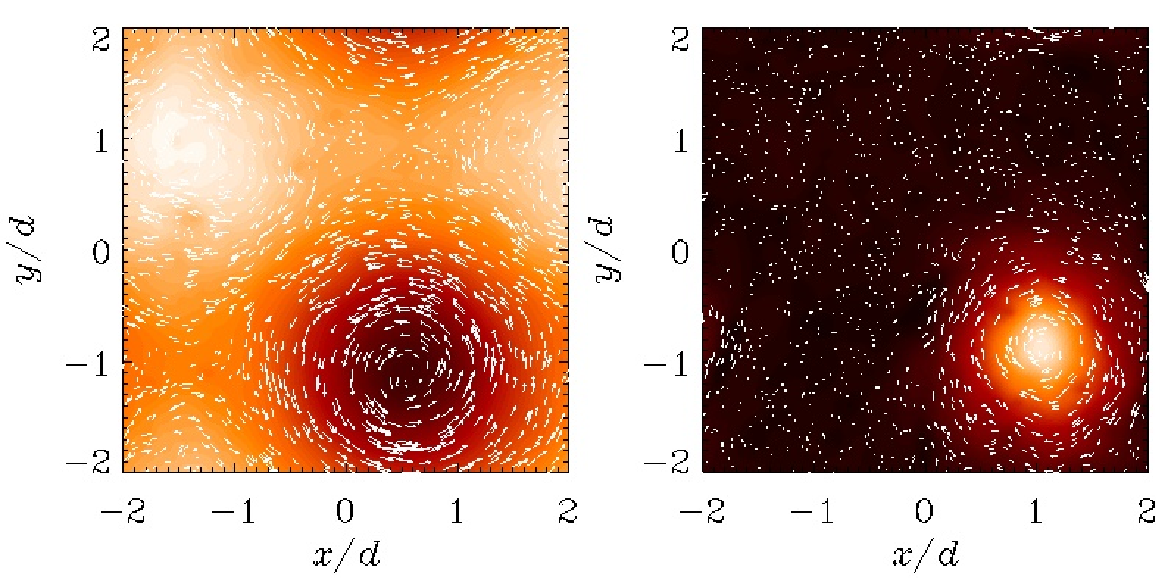}
\caption{Pressure (colors) and horizontal flows (arrows) from the 
  middle of the
  convection zone in Runs~A9 (left panel) and A1 (right panel).}
\label{pgeos}
\end{figure}

\begin{figure}[t]
\centering
\includegraphics[width=\columnwidth]{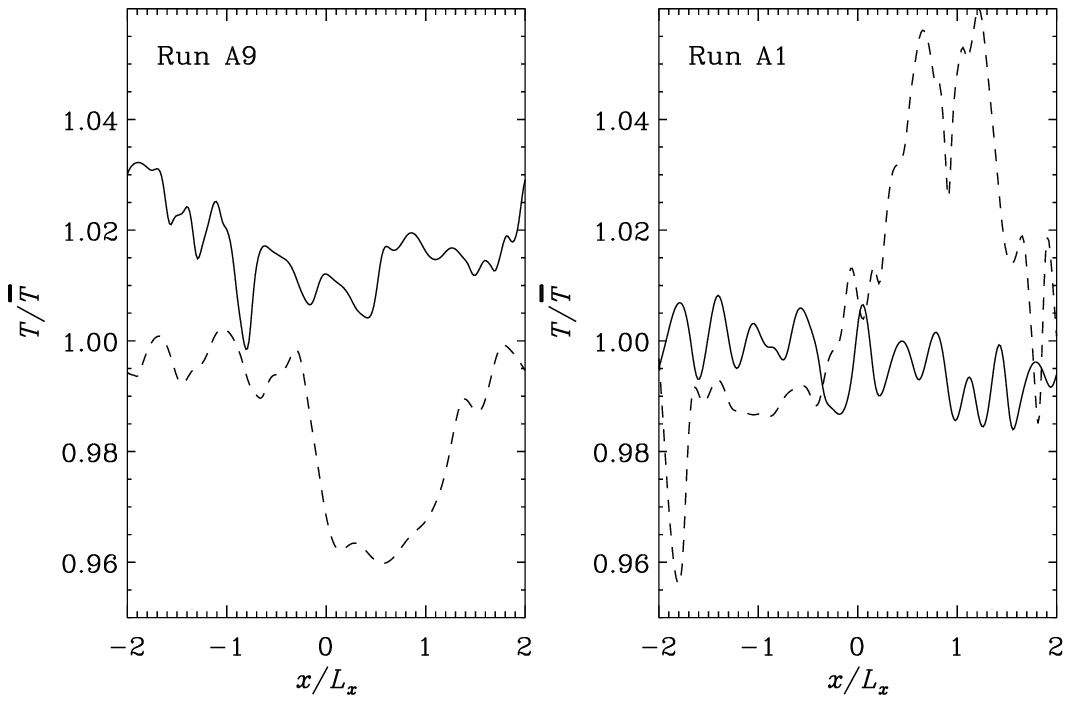}
\caption{Temperature as a function of $x$ from a quiescent (solid
  lines) and cyclonic (dashed) regions for Runs~A9 (left panel) and A1
  (right panel) from $z=z_{\rm m}$. The positions of the cuts are 
  indicated in the 
  leftmost panels of Fig.~\ref{pT_slices} with corresponding linestyles.
  The normalization factor $\overline{T}$ is the
  horizontal average of the temperature.}
\label{ptempc}
\end{figure}

Figure~\ref{pgeos} shows that in Runs~A9 and A1 the flow is in
geostrophic balance, i.e.\ that the flow follows the isocontours of
pressure for both types of vortices. The cyclone in Run~A9 
appears as a
low-pressure area, similarly to the cyclones in the atmosphere of the
Earth, whereas the anti-cyclone in Run~A1 coincides with a high
pressure region. A weaker high pressure region is present also in
Run~A9. It is not clear whether this kind of single or two spot
configuration lasts if the domain is larger in the horizontal
directions, or whether a greater number of spots appear.  We find that
the temperature contrast between the spot and the surrounding medium
is of the order of five per cent (Fig.~\ref{ptempc}) for both types of
vortices. Although the relative temperature contrast between the
vortex and the surrounding vortex-free convection seems to be a robust
feature in the simulations, we must remain cautious when comparing the
results with observations. This is due to the rather primitive nature
of the simulations that lack realistic radiation transport. Convection
in our model is also fairly inefficient by design, only 20 percent of
the total flux being carried by it.

\subsection{Dynamo considerations and discussion}
\begin{figure}[t]
\centering
\includegraphics[width=\columnwidth]{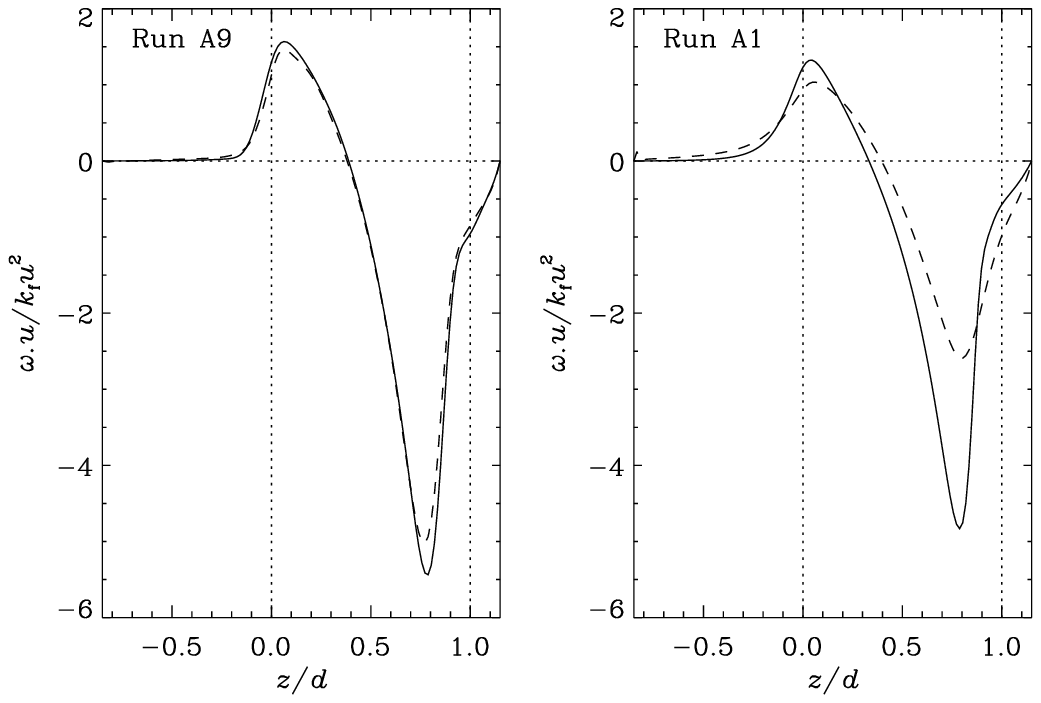}
\caption{Horizontally averaged kinetic helicity $\overline{\bm\omega\cdot
    {\bm u}}$ as a function of $z$ from a quiescent (solid lines) and
  cyclonic (dashed) states for Runs~A9 (left panel) and A1 (right
  panel). The vertical dotted lines at $z=0$ and $z=d$ indicate the
  bottom and top of the convectively unstable layer, respectively.}
\label{plot_heli}
\end{figure}

Figure~\ref{plot_heli} shows the horizontally averaged kinetic
helicity, $\overline{\bm\omega \cdot{\bm u}}$, where
$\bm{\omega}=\bm\nabla\times{\bm u}$, from Runs~A9 and A1 from the
initial, purely convective cyclone-free, and final fully saturated
stages of the simulations. The data is averaged over a period of
roughly 60 convective turnover times in each case. We find that in
Run~A9, where a cool cyclonic vortex appears, there is almost no
change in the kinetic helicity between the initial and final stages of
the simulation. In this run convection, and thus vertical motions, are
largely suppressed within the vortex (see
Fig.~\ref{puu_slices}). Furthermore, the dominant contribution to the
vorticity due to the cyclone arises via the vertical component
$\omega_z=\pd_x u_y -\pd_y u_x$, which is positive for a cyclonic
vortex. These two effects seem to compensate each other and the
helicity within the cyclone is not greatly enhanced or depressed with
respect to the surroundings. This would indicate that the influence
of the cyclonic vortices on the magnetic field amplification would be
minor, as the helicity remains unaltered. On the other hand, the
strong horizontal motions connected to the cyclone might be able to
amplify the field by advecting the field lines.

In Run~A1, on the other hand, a more pronounced effect is seen, and
the helicity is decreased up to a factor of two in the saturated stage
(see the right panel of Fig.~\ref{plot_heli}). This change is brought
about by the different handedness of the vorticity in the anti-cyclone
and by the vigorous convection within it (see the upper row of
Fig.~\ref{puu_slices}). The combination of these produces
significantly greater helicity in the anti-cyclones, but a
predominantly different sign than in the surroundings and leads to an
overall decrease noted in Fig.~\ref{plot_heli}. The decreased
helicity suggests weaker amplification of the magnetic field
by anti-cyclones compared to their surroundings. Again, the strong
large-scale horizontal motions might counteract by amplifying the field by
advection.

The simulations presented here were performed with a setup identical
to that used in \citep[][hereafter KKB09]{KKB09b} to study large-scale
dynamo (LSD) action in rotating convection. In KKB09 the generation of
large-scale magnetic fields, given that the Coriolis and magnetic
Reynolds numbers exceeded critical values, were reported. The critical
Coriolis number for LSD action was found to be roughly four, which is
close to the critical value for the cyclones to emerge.

The relation of the two phenomena is an interesting question, that can
be only partially answered by the existing magnetohydrodynamic
runs from KKB09. This is because the fluid Reynolds number in 
the runs of KKB09
was in most cases lower than $\Rey_{\rm c}$ required for the vortices
to appear. Only two runs (A10 and D1 of KKB09) are clearly in the
parameter regime exceeding the critical values found here, 
in addition to
four runs (A5, A6, B5, and C1 of KKB09) where the parameters were
close to marginal. The Reynolds and Coriolis numbers for these runs
were calculated from the saturated state of the dynamo which in all
cases lowers the turbulent velocities somewhat, decreasing 
$\Rey$ and increasing $\Co$ correspondingly.
Furthermore, a different definition of the Reynolds number was used by
KKB09 than
in the present study. A reanalysis of the data of KKB09
suggests that early stages of cyclone formation are in progress in all
of the six runs listed above. However, the magnetic field grows on a
significantly shorter timescale than the cyclones, and the magnetic
field saturates already before thousand convective turnover
times. None of the runs was continued much further than twice that,
making it impossible to decide in favor or against the maintenance of
vortices based on these runs.

Nonetheless, indications of growing cyclones appear in the kinematic
regime, i.e.\ when the magnetic field is weak in comparison to 
the
kinetic
energy of the turbulence, but they are far less clear, or even absent,
when the magnetic field saturates. This raises two related questions:
firstly, are the vortices responsible for the emergence of the
large-scale magnetic fields, and secondly, can the vortices coexist in
the regime where strong magnetic fields are present? The current data
suggests that the presence of the vortices is not essential for the
large-scale magnetic fields which persist throughout the saturated
state, whereas the vortices remain less prominent or suppressed. This
fact is related to the second issue. As noted above, the simulations
of KKB09 are too short for the vortices to fully
saturate. Thus, we cannot conclusively state whether the lack
of the vortices in the dynamo regime is due to the magnetic field
simply reducing the Reynolds number below the critical value, or a
direct influence of the Lorentz force on the growing vortices. We will
address the questions related to magnetic fields and dynamo action in
more detail in a forthcoming publication.

\subsection{Observational implications}

If large-scale cyclones such as those found in the present study occur
in real stars, they will cause observational signatures on the stellar
surface due to their lower or higher temperatures. The temperature
contrasts seen in the surface maps obtained by Doppler imaging are
somewhat stronger than the value of roughly five percent found in this
study; for instance, on the surface of the active RS CVn binary II
Peg, analyzed by \citet{Marjaana11} and \cite{Thomas11}, the coolest
spot temperatures, depending on the season, are 10-20 per cent below
the mean surface temperature. Similar spot temperatures have also been
obtained by analyzing molecular absorption bands, but cooler stars
seem to have a lower spot contrast \citep{Oneal98}. Taken that the
numerical model is quite simple, for instance in the sense that the
transport of energy by convection is underestimated, this discrepancy
is not overwhelmingly large. Interestingly, Doppler images commonly
also show hot surface features
\citep[cf.][]{Heidi07,Marjaana11,Thomas11}. These may be artefacts of
the Doppler imaging procedure, but it is not ruled out that they could
arise from real physical sources, such as the anti-cyclonic 
vortices seen in the present study.

It is obviously very hard to explain active longitudes and their
drift based on the vortex instability scenario; we believe that a
large-scale dynamo process is responsible for these basic features, as
commonly believed
\citep[e.g.][]{KR80,MBBT95,tuominen2002starspot}. Nevertheless, it is
possible that the vortex-instability contributes to the formation of
starspots, and may interfere with the dynamo-instability, especially
during the epochs of lower magnetic activity of the stellar
cycle. Although it is very hard to predict the implications of the
vortices in the magnetohydrodynamic regime, it would appear natural
that spots, either cool or warm, generated by a hydrodynamic
vortex-instability, could also contribute to the apparent
decorrelation of magnetic field from the temperature structures.

The net helicity, which is important for the amplification of the 
magnetic field, will be influenced differently by cyclones and anti-cyclones.
Anti-cyclones will decrease the net helicity, while the effect of cyclones 
will be close to zero.
This seems to imply that the magnetic field
amplification would be equally or even more difficult in the regions
of the vortices; this picture, however, may be complicated by the
presence of strong large-scale horizontal motions present in 
these structures,
that might amplify the magnetic field simply by their capability for
advecting the field lines.

\section{Conclusions}
\label{sec:conclusions}

We report the formation of large-scale vortices in rapidly rotating
turbulent convection in local f-plane simulations. The vortices appear
provided the Reynolds and Coriolis numbers exceed critical
values. Near the critical Coriolis number, the vortices are cyclonic
and cool in comparison to the surrounding atmosphere, whereas for
faster rotation warm anti-cyclonic vortices appear \citep[see
also][]{Chan07}. The relative temperature difference between the
vortex and its surroundings is of the order of five per cent in all
cases. This is of the same order of magnitude as the contrast 
deduced indirectly from
photometric and spectroscopic observations of late-type stars.
In our simulations the typical size of the vortices is comparable to
the depth of the convectively unstable layer.
However, we have not studied how the size of the structures depends
e.g.\ on the depth of the convection zone.

We propose that the vortices studied here can be present in the
atmospheres of rapidly rotating late-type stars, thus contributing to
rotationally modulated variations in the brightness and spectrum of
the star. Such features have generally been interpreted to be caused
by magnetic spots, reminiscent of sunspots. However, our results
suggest that the turbulent convection and rapid rotation of these
stars can generate large-scale temperature anomalies in their
atmospheres via a purely hydrodynamical process. Similar
vortex-structures are observed in the atmospheres of Jupiter and
Saturn. Although their definitive explanation is still debated, it is
possible that they are related to rapidly rotating thermal convection
in their atmospheres.

However, several issues remain to be sorted out before the reality of
cyclones and anti-cyclones in the surface layers of stars can be
established. The current model is highly simplified and neglects the
effects of sphericity and magnetic fields. In spherical geometry more
realistic large-scale flows can occur which might lead to other
hydrodynamical instabilities. However, current rapidly rotating
simulations in spherical coordinates have not shown evidence of
large-scale vortices \citep[e.g.][]{BBBMT08,KKBMT10,KMGBC11}, although
non-axisymmetric features are seen near the equator \citep{BBBMT08}.
It is possible that the lack of large-scale vortices in these
simulations is related either to the lack of spatial resolution or too
short integration time.

Magnetic fields, on the other hand, are ubiquitous in stars with
convection zones. Furthermore, on the Sun they form strong flux
concentrations, i.e.\ sunspots. At the moment, direct simulations
cannot self-consistently produce sunspot-like structures in local
geometry \citep[e.g.][]{KBKMR11}. However, the magnetic fields in
global simulations are also very different from the high-latitude
spots and
active longitudes deduced from observations, namely showing more
axisymmetric fields residing also near the equator
\citep[e.g.][]{KKBMT10,BMBBT11}. The apparently poor correlation
between magnetic fields and temperature anomalies in surface maps
based on Doppler imaging also suggests that an alternative mechanism
might be involved. The presence of large-scale high-latitude vortices
presents such an alternative.

Currently it is not clear what happens to the vortices when magnetic
fields are present. Our previous dynamo simulations in the same
parameter regime \citep{KKB09b} did not show clear signs of vortices
in the saturated regime of the dynamo although this might be explained
by the too short integration time. Addressing this issue, however, is
not within the scope of the present paper and we will revisit it in a
future publication.

\acknowledgements
  The authors thank the anonymous referee on constructive comments
  on the paper.
  The simulations were performed using the supercomputers hosted by CSC
  -- IT Center for Science Ltd.\ in Espoo, Finland, who are administered
  by the Finnish Ministry of Education.  Financial support from the
  Academy of Finland grants No.\ 136189, 140970 (PJK) and 218159, 141017
  (MJM), and the `Active Suns' research project at University of
  Helsinki (TH) is acknowledged. The authors acknowledge the hospitality
  of NORDITA during their visits.

\bibliography{paper}

\end{document}